\documentclass[12pt]{article}

\newcommand{\sect}[1]{\setcounter{equation}{0}\section{#1}}

\newcommand{\eq}{\begin{equation}}
\newcommand{\eqa}{\begin{eqnarray}}
\newcommand{\en}{\end{equation}}
\newcommand{\ena}{\end{eqnarray}}
\newcommand{\enn}{\nonumber \end{equation}}

\def\epsihat{{\widehat\epsilon}}
\def\epsilonhat{{\widehat{\epsilon}}}
\def\deltahat{ {\widehat\delta} }
\def\Omhat{\widehat{\Om}}
\def\Vhat{\widehat{V}}

\def\Vtildehat{{\widehat{\Vtilde}}}
\def\Rtildehat{{\widehat{\Rtilde}}}

\def\phihat{\widehat{\phi}}
\def\dehat{\widehat{\de}}

\def\epsitildehat{\widehat{\epsitilde}}

\def\sk{\vskip .4cm}
\def\noi{\noindent}
\def\om{\omega}
\def\al{\alpha}

\def\ga{\gamma}
\def\Ga{\Gamma}

\let \part\partial

\def\unquarto{{1 \over 4}}

\def\unmezzo{{1 \over 2}}
\def\epsi{\varepsilon}
\def\we{\wedge}

\def\de{\delta}

\def\part{\partial}

\def\sk{\vskip .4cm}

\def\noi{\noindent}

\def\X0{X^0}

\def\om{\omega}

\def\al{\alpha}
\def\ga{\gamma}

\def\unquarto{{1 \over 4}}
\def\unmezzo{{1 \over 2}}
\def\epsi{\varepsilon}

\def\we{\wedge}

\def\de{\delta}

\def\Rhat#1#2{ \Rh^{#1}_{~~~#2} }

\def\square{{\,\lower0.9pt\vbox{\hrule \hbox{\vrule height 0.2 cm
\hskip 0.2 cm \vrule height 0.2 cm}\hrule}\,}}

\def\westar{\we_\star}

\def\omtilde{\tilde \om}
\def\Vtilde{\widetilde{V}}
\def\Ttilde{\tilde T}
\def\Rtilde{\widetilde{R}}

\def\epsitilde{\widetilde{\epsi}}

\def\Om{\Omega}

\def\phitilde{\tilde \phi}
\def\varphitilde{\tilde \varphi}
\def\Rhat{\widehat{R}}
\def\omhat{\widehat{\om}}
\def\omtildehat{\widehat{\omtilde}}
\def\phihat{\widehat{\phi}}
\def\Phihat{\widehat{\Phi}}
\def\varphihat{\widehat{\varphi}}

\def\phitildehat{{\widehat{\phitilde}}}
\def\varphitildehat{{\widehat{\varphitilde}}}
\def\nn{\nonumber}

\begin{document}

\begin{titlepage}
\begin{center}{\Large \bf Extended gravity theories  from dynamical noncommutativity}
\\[3em]
{\large {\bf Paolo Aschieri} and {\bf Leonardo Castellani}} \\ [2em] {\sl Dipartimento di Scienze e Innovazione Tecnologica
\\ INFN Gruppo collegato di Alessandria,\\Universit\`a del Piemonte Orientale,\\ Viale T. Michel 11,  15121 Alessandria, Italy}\\ [1.5em]
\end{center}

\begin{abstract}

\vskip 0.2cm
In this paper we  couple noncommutative (NC) vielbein
gravity to scalar fields. Noncommutativity is encoded in a
$\star$-product between forms, 
given by an abelian twist (a twist with commuting vector fields).   A
geometric generalization of the Seiberg-Witten map for abelian twists
yields an extended theory of gravity coupled to scalars, where all
fields are ordinary (commutative) fields. The vectors defining the
twist can be related to the scalar fields and their derivatives, and
hence acquire dynamics. Higher derivative corrections to the classical
Einstein-Hilbert and Klein-Gordon actions  are organized in successive powers of the noncommutativity parameter $\theta^{AB}$.

 \end{abstract}

\vskip 10cm \noi \hrule \vskip.2cm \noi {\small aschieri@to.infn.it\\
leonardo.castellani@mfn.unipmn.it }

\end{titlepage}

\newpage
\setcounter{page}{1}

\sect{Introduction}

In this paper we study the coupling of scalar fields to the
noncommutative gravity theory constructed in \cite{AC1} and further
developed in \cite{AC3,AC4}. A noncommutative action is found, and
generalizes the classical Einstein-Hilbert + Klein-Gordon actions. It
is invariant under diffeomorphims and noncommutative local Lorentz
transformations. The noncommutativity is governed by an abelian twist,
${\cal F}=e^{-\frac{i}{2}\theta^{AB}X_A\otimes X_B}$, and the corresponding $\star$-product between forms reads:
    \eqa
    \tau \westar \tau' &\equiv & \sum_{n=0}^\infty \left({i \over 2}\right)^n \theta^{A_1B_1} \cdots \theta^{A_nB_n}
   (\ell_{X_{A_1}} \cdots \ell_{X_{A_n}} \tau) \we  (\ell_{X_{B_1}} \cdots \ell_{X_{B_n}} \tau')  \nonumber \\
  = &\!\!\!\!\!\!&\!\!\!\!\!\!\!\!\!\!\!\!\tau \we \tau' + {i \over 2} \theta^{AB} (\ell_{X_A} \tau) \we (\ell_{X_B} \tau') + {1 \over 2!}  {\left( i \over 2 \right)^2} \theta^{A_1B_1} \theta^{A_2B_2}  (\ell_{X_{A_1}} \ell_{X_{A_2}} \tau) \we
 (\ell_{X_{B_1}} \ell_{X_{B_2}} \tau') + \cdots \nonumber 
  \label{defwestar}
  \ena
       \noi where the mutually commuting vector fields $X_A$ act on forms via the Lie derivatives  ${\ell}_{X_A} $. 
     This product is associative, and the above formula holds also for
     $\tau$ or $\tau'$ being  $0$-forms (i.e.
     functions)\footnote{when restricted to $0$-forms, and if $X_A=
       \de^\mu_A \,{\part \over \part x^\mu}$, the $\star$-product
       reduces to the well-known Moyal-Groenewold product
       \cite{MoyalGroenewold}.}. A different study of a Klein-Gordon
     action in a curved background is presented in
\cite{Ohl:2009qe, Schenkel:2010sc}, 
     based on the metric formulation of twist noncommutative gravity
     \cite{Wess}, where noncommutative local Lorentz symmetry is absent.

Use of the geometric generalization \cite{AC3} of the Seiberg-Witten
map \cite{SW} between noncommutative and commutative local Lorentz symmetry allows to reinterpret the noncommutative vielbein gravity coupled to scalar fields as a theory with ordinary fields on commutative spacetime, invariant under diffeomorphisms and usual local Lorentz rotations. It is a particular higher derivative extension of Einstein gravity coupled to scalar fields.

The commuting vectors $X_A$ present in the twist  also enter the action, but they can be related to the scalar fields, so that the resulting theory contains only the vierbein, the spin connection and the scalars.  Alternatively one can 
keep the vectors $X_A$ as independent fields, and introduce a corresponding kinetic term coupled to the gravity action.

In the first scheme the particular extension of Einstein gravity depends on how the vectors $X_A$ are related to the scalars. This relation is controlled by a function $Z$ of the scalars, a sort of ``potential" for noncommutativity. The $X_A$ vector fields are given in terms of (derivatives of) $Z$, and therefore are not anymore background spectators but acquire dynamics induced by the scalars. Some choices for $Z$ are discussed. 
\sk
The paper is organized as follows. In Section 2 we recall the
geometric action for scalars coupled to gravity, and show how it can
be recast in an index-free form, suitable for a noncommutative
generalization. In Section 3 the noncommutative action is obtained,
and its symmetries are discussed. In Section 4 we discuss the noncommutative field equations and how to find their solutions. We show that the nondynamical fields of the classical theory remain nondynamical in the NC theory. The geometric Seiberg-Witten map for
abelian twists is recalled in Section 5, and applied in Section 6 to
show that the first order correction (in $\theta$) of the action
vanishes. In Section 7 we give the second order expansion of the
scalar action, in a manifestly Lorentz gauge invariant form. In
Section 8 we relate the $X_A$ vector fields to the scalars and
discuss the  potential $Z$. An Appendix summarizes the $D=4$ gamma matrix conventions.

\sect{Scalars coupled to gravity: classical action}

\subsection{Geometric first order action}

The classical action for a scalar multiplet $\phi^I$ coupled to gravity can be written in first order form as follows:
\eq
 S= \int \Big(  R^{ab} \we V^c \we V^d + {1\over 3} \varphi^{Ia} d\phi^I \we V^b \we V^c \we V^d - {1 \over 4!} (\varphi^{Ir} \varphi^I_r + W(\phi^I)) V^a \we V^b \we V^c \we V^d \Big) \epsi_{abcd} \label{scalarsplusgravity}
 \en

\noi The fundamental fields are:
 \sk
 \noi i) the one-form spin connection $\om^{ab}$, entering the action through the
Lorentz curvature
 \eq
  R^{ab} \equiv d\om^{ab} - \om^{ac} \we \om^{cb}
  \en
 
\noi ii) a multiplet of $N$ scalar fields (zero-forms) $\phi^I$, ($I=1,...N$)

\noi  iii) the zero-form auxiliary  fields $\varphi^I_a$ ;

 \noi iv) the vielbein one-form $V^a$
\sk
 The variation of the action with respect to $\varphi^I_a$
identifies the auxiliary field with the derivative of the scalar field:
 \eq
  \varphi^I_{a} = \part_a \phi^I \label{auxeqclass}
 \en
 \noi where $\partial_a=V^\mu_a\partial_\mu$, $V^\mu_a$ being the
 inverse matrix of $V^a_\mu$, with $V^a=V^a_\mu dx^\mu$.  The field equation for $\om^{ab}$ gives the zero torsion condition:
  \eq
   d V^a - \om^{ab} \we V^b =0  \label{omeqclass}
    \en
     which allows to express the spin connection in terms of derivatives of vielbeins and inverse
     vielbeins (second order formalism).
     
     The variation with respect to the scalar fields $\phi^I$
 yields the Klein-Gordon equation in curved space:
 \eq
  D_a \part^a \phi^I + 12~ {\de W  \over \de \phi^I} =0
 \en
where $W$ is the scalar potential, depending only on $\phi^I$, $D_a
=$ Lorentz covariant derivative,  and where repeated indices are summed with the
flat Minkowski metric $\eta_{ab}$.

Finally the variation of the vielbein yields the Einstein equations:
\eq
 R^{ra}_{~~rb} - \unmezzo \de^a_b R = -\unquarto (\part^a \phi^I \part_b \phi^I - \unmezzo \de^a_b \part^r \phi^I \part_r \phi^I ) - 3 W \de^a_b
 \en
\sk
\noi
{\bf Note:} the use of auxiliary fields $\varphi^{Ia}$ in the action
(\ref{scalarsplusgravity}) allows to avoid the appearance of  the
Hodge star operation. cf. also the action for NC Yang-Mills coupled to
gravity in ref. \cite{AC4}. This formulation is useful since the
noncommutative generalization of the Hodge star operation in case of
an arbitrary 
curved metric is presently an open question. The introduction of the
auxiliary fields bypasses this problem, and still leads in the classical limit to usual Einstein gravity coupled to scalars. 

\subsection{Index-free action}

 The action (\ref{scalarsplusgravity}) can be recast in index-free form as follows:
\eq 
 S =  \int Tr \Big( i \ga_5 (R \we V \we V - {1 \over 4!} \varphi^I \varphi^I V \we V \we V \we V + {1 \over 3} \varphi^I D\Phi^I \we V \we V \we V ) \Big)
 \label{scalarindexfree}
\en
\noi where $R \equiv \unquarto R^{ab} \ga_{ab}$, 
 $\Phi^I \equiv \phi^I 1$, $\varphi^I \equiv \varphi^{Ia} \ga_a$, $V \equiv V^a \ga_a$,
and the trace is taken  on the
spinor space. We take for simplicity $W=0$. 
Use of the $D=4$ gamma matrix identities
 \eq
  Tr (\ga_{ab} \ga_c \ga_d \ga_5 ) = -4 i \epsi_{abcd}, ~~~ \ga_{[a} \ga_b \ga_c \ga_{d]} = -i \ga_5 \epsi_{abcd}
   \en
   yields the action (\ref{scalarsplusgravity}). The index-free definition of the Lorentz curvature is
 \eq
 R = d \Om - \Om \we \Om \label{defR}
 \en
\noi with $\Om \equiv \unquarto \om^{ab} \ga_{ab}$. The definition  (\ref{defR}) implies the
Bianchi identities for $R$:
\eq 
 D R \equiv d R - \Om \we R + R \we \Om =0
 \en
For notational economy, in the following we will occasionally omit the multiplet indices $I$ in the scalar fields.

\subsection{Symmetries}

Apart from general coordinate invariance, obtained {\it ab initio}
through the use of Cartan calculus, the action (\ref{scalarsplusgravity}) is invariant under local Lorentz transformations. In index-free form these transformations are
  \eqa
 & & \de_\epsi \Phi= 0,~~ \de_\epsi \varphi= 0\\
  & &\de_\epsi V =  - V \epsi + \epsi V \label{LorV}\\
 & &  \de_\epsi \Om =  d \epsi -\Om  \epsi + \epsi  \Om ~~~\Rightarrow
\de_\epsi  R= -R \epsi + \epsi R \label{Lorom}
   \ena
\noi with 
 $ \epsi \equiv \unquarto \epsi^{ab} \ga_{ab}$. Invariance of the action 
 (\ref{scalarindexfree}) under these transformations  immediately
 follows from the cyclicity of the trace and the fact that $\epsi$ 
commutes with $\ga_5$.
 
\sect{Noncommutative action}

The twisted noncommutative action is found by replacing in the 
index-free action (\ref{scalarindexfree})  all products by twisted
$\star$-products:
\eq
  S =  \int Tr \Big( i \ga_5 (\Rhat \westar \Vhat \westar \Vhat - {1 \over 4!} \varphihat \star \varphihat \star \Vhat \westar \Vhat \westar \Vhat \westar \Vhat + {1 \over 6} (\varphihat \star D\Phihat + D\Phihat \star \varphihat) \westar \Vhat \westar \Vhat \westar \Vhat ) \Big)
 \label{scalartwisted}
\en
\noi where the covariant exterior derivative is defined by:
 \eq
  D \Phihat \equiv d \Phihat - \Omhat \star \Phihat + \Phihat \star \Omhat
   \en
  and  the  curvature $\Rhat (\Omhat)$ is 
 \eq
\Rhat (\Omhat) = d \Omhat- \Omhat \westar \Omhat
 \en
where $\widehat{\Omega}$ is the noncommutative spin connection matrix. This definition implies the Bianchi identity:
 \eq
 D \Rhat(\Omhat) \equiv  d \Rhat(\Omhat) - \Omhat \westar \Rhat(\Omhat) +  \Rhat(\Omhat)  \westar \Omhat =0
 \en

The (associative) $\star$-exterior product between forms is defined by using Lie derivatives along a set
of commuting vector fields $X_A$ (see the formula given in the Introduction. For a summary on twist differential geometry see for ex. the Appendix of ref. \cite{AC1}). The symmetrization in the third term of the noncommutative action (\ref{scalartwisted})
is necessary for the action to be real. 

The NC fields have deformed transformation laws: to distinguish them from the ordinary fields transforming under the usual laws we denote them with a hat. In fact, the Seiberg-Witten map 
relates the hatted fields (the``noncommutative" fields) to the ordinary ones.

\subsection{Noncommutative symmetries}

 The NC action (\ref{scalartwisted}) is invariant under general coordinate transformations (being the integral of a 
 4-form) and under the $\star$-gauge variations:
\eqa\
 {\deltahat_{\hat\epsilon}}
 \Omhat&=&d \epsihat - \Omhat \star \epsihat + \epsihat \star \Omhat ~~\Longrightarrow~~
{\deltahat_{\hat\epsi}}
 \Rhat(\Omhat) = -\Rhat(\Omhat)  \star \epsihat + \epsihat \star
 \Rhat(\Omhat) \nn\\
  \deltahat_{\hat\epsilon} \Vhat &=& -\Vhat \star \epsihat + \epsihat \star
  \Vhat  \nn\\
 \deltahat_{\hat\epsilon} \varphihat &=& - \varphihat \star \epsihat + \epsihat \star \varphihat\nn\\
  \deltahat_{\hat\epsilon} \Phihat &=& - \Phihat \star \epsihat + \epsihat \star \Phihat  ~~\Longrightarrow~~
{\deltahat_{\hat\epsilon}}
(D \Phihat )= -(D \Phihat ) \star \epsihat + \epsihat \star
(D \Phihat )\label{*gauge}
   \ena
   with an arbitrary parameter $\epsihat$ commuting with $\ga_5$. 

\noi The invariance of the noncommutative action under these transformations
relies on the cyclicity of the integral (and of the trace) and on
$\epsihat$ commuting with $\ga_5$.

Because of noncommutativity, extra fields are entering in
the expansions of $\Omhat, \Vhat, \Phihat, \varphihat$.  

 \noi Indeed now the $\star$-gauge variations of the fields (\ref{*gauge}) include also anticommutators of gamma matrices,
 due to the noncommutativity of the $\star$-product. Since for example the anticommutator
 $\{ \ga_{ab},\ga_{cd} \}$ contains $1$ and $\ga_5$, we see that the corresponding fields
 must be included in the expansion of $\widehat\Om$. Similarly,
 $\widehat V$ must contain a $\ga_a \ga_5$ term due
 to $\{ \ga_{ab},\ga_{c} \}$, etc. The composition law for gauge parameters becomes:
 \eq
   [\de_{\hat\epsilon_1},\de_{\hat\epsilon_2}] = \de_{\hat\epsilon_2 \star \hat\epsilon_1 -
   \hat\epsilon_1 \star \hat\epsilon_2 }
   \en
   \noi so that $\epsilon$ must contain the $1$ and $\ga_5$ terms, since they appear in the
   composite parameter $\epsihat_2 \star \epsihat_1 - \epsihat_1 \star \epsihat_2$.

The $SO(1,3)$ enveloping algebra generators  $\Ga_\al$
 are chosen to be $\ga_0$-antihermitian:
 \eqa
 & & \Ga_\al  = {1\over 4} \ga_{ab},~i1,~\ga_5 \\
 & & (\Ga_\al)^\dagger = - \ga_0 \Ga_\al \ga_0
\ena

The noncommutative fields are expanded as follows:
 \eqa
 & & \Omhat = {1 \over 4} \omhat^{ab}  \ga_{ab}  + i \omhat 1 + \omtildehat \ga_5  \label{Omhat}\\
 & & \Vhat =  \Vhat^{a} \ga_a  + \Vtildehat{}^a \ga_a \ga_5 \\
 & & \Phihat^I =i {1\over 4} \phihat^{I~ab} \ga_{ab} +   \phihat^I 1+ 
 i \phitildehat{}^I  \ga_5 \label{phihat}\\
& &  \varphihat^I = \varphihat^{Ia} \ga_{a} +   \varphitildehat{}^{Ia} \ga_a \ga_5 \label{varphihat}
 \ena

Similarly, for the curvature and the gauge parameter the expansions are: 
 \eqa
& &
 \Rhat= {1 \over 4} \Rhat^{ab}  \ga_{ab}  +i_{\,} \widehat r  +
\Rtildehat \ga_5  \\
& & \epsilonhat =  {1 \over 4} \widehat\varepsilon^{\,ab}  \ga_{ab}  +  i_{\,} \widehat\varepsilon + \epsitildehat \ga_5 
\ena
 All the components along the $SO(1,3)$ enveloping algebra generators are taken to be real, and therefore fields and curvatures satisfy the hermiticity properties: 
 \eq
 \Omhat^\dagger = - \ga_0 \Omhat \ga_0,~\Vhat^\dagger = \ga_0 \Vhat \ga_0,~\Phihat{}^\dagger = \ga_0 \Phihat \ga_0,  ~\varphihat{}^\dagger = \ga_0 \varphihat \ga_0,~\Rhat^\dagger =- \ga_0 \Rhat \ga_0
 \en 
\noi i.e. $\Omhat$ and $\Rhat$ are $\ga_0$-antihermitian, while $\Vhat$, $\varphihat$ and $\Phihat$ are $\ga_0$-hermitian. Using these rules it is a quick matter to check that the noncommutative action
(\ref{scalartwisted}) is real. 

\sk

The NC action is also invariant under charge conjugation, defined on  elementary
fields as :
\eq
\widehat V\to \widehat V^C=C\widehat V^TC~,~~\widehat \Omega\to \widehat \Omega^C=C\widehat \Omega^TC~,~~
\widehat \Phi\to \widehat \Phi^C=-C\widehat \Phi^TC~,~~\widehat \varphi\to\widehat \varphi^C=C\widehat \varphi^TC~.
\en
Charge conjugation is extended linearly and antimultiplicatively
on
$\star$-products (but not on matrix products) of fields, so that
$(f\star g)^C=g^C\star f^C$, i.e.,
\eq
(f\star g)^C=f^C\star_{-\theta}g^C \label{ccff}
 \en
 where $\star_{-\theta}$ denotes the star product with opposite
noncommutative parameters $-\theta^{AB}$. This formula holds also
 for matrix valued
fields.

The three addends in the action (\ref{scalartwisted}) are separately
charge conjugation invariant. Charge conjugation symmetry of the second
addend is easily verified:
\eqa
\!\!\!\!\!\Big(\int Tr (i \ga_5  \varphihat \star \varphihat \,\star &&\!\!\!\!\!\!\!\!\!\!\!\!\!\!\!\!\!\!\!\Vhat \westar
\Vhat \westar \Vhat \westar \Vhat) \,\Big)^C=\nn\\
&~~=& \int Tr (i \ga_5  \varphihat^C \star_{-\theta} \varphihat^C \star_{-\theta} \Vhat^C {\westar}_{{_{\!-\theta}}}
\Vhat^C {\westar}_{{_{\!-\theta}}} \Vhat^C {\westar}_{{_{\!-\theta}}} \Vhat^C)\nn\\
&~~=&
\int Tr (C(-i) \ga_5  C\,\varphihat^T \star_{-\theta} \varphihat^T \star_{-\theta} \Vhat^T {\westar} _{{_{\!-\theta}}}
\Vhat^T {\westar}_{{_{\!-\theta}}} \Vhat^T {\westar}_{{_{\!-\theta}}}
\Vhat^T)\nn\\
&~~=&
\int Tr (i \ga^T_5  \varphihat^T \star_{-\theta} \varphihat^T \star_{-\theta} \Vhat^T {\westar}_{{_{\!-\theta}}}
\Vhat^T {\westar}_{{_{\!-\theta}}} \Vhat^T {\westar}_{{_{\!-\theta}}}
\Vhat^T)\nn\\
&~~=&
\int Tr (\Vhat\westar
\Vhat\westar\Vhat \westar
\Vhat\star \varphihat \star \varphihat\, i\gamma_5)\nn\\
&~~=&
\int Tr (i \ga_5  \varphihat \star \varphihat \,\star \Vhat \westar
\Vhat \westar \Vhat \westar \Vhat)~.
\ena
One proceeds similarly for the first addend, i.e. the pure NC gravity
term (an explicit proof is in \cite{AC3}). For the last term one uses 
 the definition of charge conjugation on the scalar 
$\widehat\Phi$ and the consequent property
$(D\widehat\Phi)^C=-C(D\widehat\Phi)^T C$.

\sk
The noncommutative fields can be considered as dependent on $\theta$, since 
their $\star$-gauge transformed images are $\theta$ dependent. Indeed 
 the $\star$-gauge transformations depend on $\theta$ (since the $\star$-product depends on $\theta$), so that  the transformed fields necessarily depend on $\theta$.
We can then expand the fields in power series of the noncommutativity
parameter $\theta$, each coefficient of a given power of $\theta$
being a new field. These infinite degrees of freedom can be reduced by
requiring the fields to satisfy the constraints
\eq
\widehat V^C=\widehat V_{-\theta}~,~~\widehat\Omega^C=\widehat\Omega_{-\theta}~,~~\widehat\Phi^C=\widehat\Phi_{-\theta}~,
~~\widehat\varphi^C=\widehat\varphi_{-\theta}~.
\label{ccc}\en
These conditions are compatible with the $\star$-gauge variations
(\ref{*gauge}) provided that $\widehat{\epsilon}^C=\widehat{\epsilon}_{-\theta}$. 
They are equivalent to require that the component fields
$\widehat\omega^{ab}$, $\widehat V^a$, $\widehat\phi^I$,
$\widehat\varphi^{I\,a}$, $\widehat  \epsi^{ab}$ are even in $\theta$ while the 
other components are odd in $\theta$.

We can further constrain these fields so that
the noncommutative theory is reduced to a theory with  the same degrees of
freedom of the classical theory. This is done in Section 5 via the Seiberg-Witten map, that allows  
 to express all noncommutative fields in terms of the 
commutative -or classical- fields $V^a,~\omega^{ab},~ \phi^I,~ \varphi^{Ia}~.$
\sk

To conclude this section, we remark that conditions (\ref{ccc}) imply that the NC action {\it must be even in} $\theta$.
Indeed, because of  (\ref{ccc}) and (\ref{ccff}), $S_ {\theta} $ is mapped into
$S_{-\theta}$ under charge conjugation. Invariance of the action under charge conjugation then implies invariance of
the action under $\theta\to -\theta$.
Finally $S_\theta=S_{-\theta}$ implies that all corrections to the
classical action are even in $\theta$.

\sect{NC field equations and perturbative solutions}

In this section $\star$-products, $\westar$ products and hats 
$\widehat{~\,~}$  are omitted for notational brevity.
The NC field equations, obtained by varying the NC action (\ref{scalartwisted}), are given by:
\sk
\noi {\bf Auxiliary field $\varphi$:}
\eq
 Tr \left(  \Ga_{a,a5} ([VVVV,\varphi] + 4 \{ VVV, D \Phi \} ) \right) =0
 \en
\noi {\bf Spin connection  $\Om$:}
 \eq
 Tr \left(  \Ga_{ab,1,5} (  [T,V] + {1 \over 6} ([\Phi,VV\varphi]  - [\Phi, \varphi VVV] ) \right) =0
\label{TVeq} \en
{\bf Scalar field $\Phi$: }
\eq
Tr \left(  \Ga_{ab,1,5} (- \{VVV,D \varphi \} + [ \varphi , TVV - VTV  + VVT ] ) \right) =0
 \en
{\bf Vielbein $V$: }
\eq
 Tr \left( \Ga_{a,a5} ( - \{ V,R \} + {1 \over 4!} (\{ VVV, \varphi\varphi \} +  \{V,V\varphi\varphi V \} ) - {1 \over 6} ( \{ VV, \{ \varphi, D \Phi \} \} - V 
 \{ \varphi, D \Phi \} V ) )  \right)=0
 \en
 \noi where   $\Ga_{a,a5}$  indicates $\ga_{a}$ and $\ga_a \ga_5$
  (thus there are two distinct equations) and likewise for
 $\Ga_{ab,1,5}$ (three equations corresponding to $\ga_{ab}$, $1$ and
 $\ga_5$). The noncommutative torsion two-form $T$ is defined by:
  \eq
  T \equiv T^a \ga_a + \Ttilde^a \ga_a \ga_5 \equiv  DV \equiv dV -\Om V - V \Om
  \en
 The NC field equations can be expanded in $\theta$, and have the general structure:
\eq
 E_{0} (\phi) +  E_{1} (\phi) +  E_{2}(\phi) + \cdots =0 \label{NCFE}
 \en
\noi where $\phi$ are the fields appearing in the action and
$E_0$ is the classical field equation, $E_1$ is
linear in $\theta$, $E_2$ is quadratic in $\theta$, etc. The solutions of these NC equations
will in general depend on $\theta$:
 \eq
 \phi = \phi_{0}+ \phi_{1} +  \phi_{2}+ \cdots \label{NCsol}
 \en
where $\phi_0$ is the classical field, $\phi_1$ is linear in $\theta$, $\phi_2$ is quadratic in
$\theta$, etc. Substituting the expansion 
(\ref{NCsol}) into the NC equations (\ref{NCFE}), and requiring that the coefficients
of all powers of $\theta$ vanish, we find 
 \eqa
 & & E_{0}(\phi_{0})=0 \nn\\
 & & E_{0} (\phi_{1},\phi_{0}) + E_{1}(\phi_{0}) =0 \nn\\
 & & E_{0} (\phi_{2},\phi_{1},\phi_{0}) + E_{1}(\phi_{1},\phi_{0}) + E_{2}(\phi_{0})=0 \nn\\
 & & \cdots \label{eqlist}
 \ena
The zero-th order solution (the "classical solution") of the first equation can be substituted in the second equation, which then determines the first order correction $\phi_{1}$ in terms of $\phi_{0}$. Inserting $\phi_{1}$ into the third equation enables to find $\phi_{2}$ and so on: in this way the solution of the NC field equations can be constructed order by order.

Let us see how it works in our specific example of scalar fields coupled to NC gravity. We will not try here to solve the NC field equations for the dynamical fields $V$ and $\Phi$, but will show that the nondynamical fields of the classical theory, i.e. the spin connection $\Om$ and the auxiliary field $\varphi$, remain nondynamical also in the NC context. 

This can be understood as follows.The zero-th order field equations for $\varphi$ and $\Om$ are just the classical ones given in (\ref{omeqclass}) and (\ref{auxeqclass}), allowing to express the classical spin connection in terms of the classical vielbein and the classical auxiliary field as derivative of the classical scalar field.  This is because the classical $\Om$ and $\varphi$ appear algebraically in the classical field equations.  For example $\varphi$ appears in the classical equation as $ (\varphi_{0})_{a}^{I} \epsi_{bcde} V_{0}^{b} V_{0}^{c} V_{0}^{d} V_{0}^{e}$ (and is equated to $\part_{a} \Phi_{0}^{I} \epsi_{bcde} V_{0}^{b} V_{0}^{c} V_{0}^{d} V_{0}^{e}$). The higher order corrections $\Om_{i}$, $\varphi_{i}$ {$i \ge 1$}
are determined by the higher order equations in (\ref{eqlist}), where they appear algebraically exactly as in the classical case: indeed
the first order field equation contains $\varphi_{1}$ as $(\varphi_{1})_{a}^{I}\epsi_{bcde} V_{0}^{b} V_{0}^{c} V_{0}^{d} V_{0}^{e} $ and similarly for higher orders. Then all the higher order field equations can be solved algebraically in the same way for all the $\varphi_{i}$.
The same occurs for the higher order corrections of the spin
connection that appears algebraically in the torsion field and hence
algebraically in (\ref{TVeq}). Thus we can use the NC field equations
to eliminate $\varphi$ and $\Om$, i.e. the transition from first to
second order formalism is possible also in the NC theory.
\sk
Finally, the same conclusion holds when use is made of the
Seiberg-Witten map between noncommutative and commutative local Lorentz symmetry.
In this case the fields are given from the start
(off shell) a precise $\theta$ dependence (in terms of the classical fields)
dictated by the SW map.
Substituting their expansion in the
action, after expanding also the $\star$ products, one obtains an
extended action $S_{0} + S_{1} + S_{2} +  \cdots$ expanded in powers
of $\theta$. Consider now the field equations: 
they involve only the classical fields and their higher derivatives, and can 
be expanded in powers of $\theta$. We can look for
perturbative solutions given by fields expanded in powers of $\theta$.
The reasoning is now identical to the
one used in the previous paragraph. Since at zero-th order in $\theta$
the auxiliary field $\varphi$ and the spin connection $\Omega$ are
nondynamical, by the same argument the higher order corrections to these fields
will also be nondynamical.

 \sect{Geometric Seiberg-Witten map and fields at first order in $\theta$ for a general abelian twist}

As shown in ref. \cite{AC3}, the SW map can be recast in a
coordinate-independent form, and generalized to a $\star$-product
originating from an arbitrary abelian twist. 
We expand the noncommutative fields in addends of  homogeneous degree in $\theta$,
\eqa
& &\widehat\Omega=\Omega+\Omega^1+\Omega^2+\ldots\Omega^n+\ldots\\
& &\widehat\epsilon=\epsi+\epsi^1+\epsi^2+\ldots\epsi^n+\ldots\\
& &\widehat\phi=\phi+\phi^1+\phi^2+\ldots \phi^n+\ldots
\ena
where $\Omega, \epsilon $ are the classical gauge
potential and gauge parameter, and $\phi$ is a classical scalar field, while
the fields $\Omega^n$, $\epsi^n$ and $\phi^n$  are homogeneous of order
$n$ in powers of the noncommutativity  $\theta$ and depend also from
the derivatives of the gauge potential $\Omega$, and, in case of the
gauge parameter and  the scalar field, also on the
classical field $\epsilon$, $\phi$ and their derivatives, respectively.

We recall the relevant formulae for the recursive relations
determining the SW \cite{AC3}, 
\eqa
& & 
\Omega^{n+1}=\frac{i}{4(n+1)}\theta^{AB}\{\widehat\Omega_A, 
\ell_B \widehat\Omega+ \widehat R_{B} \}^n_\star \label{rec1}\\
& & \epsi^{n+1}=  {i \over 4(n+1)} \theta^{AB} \{\Omhat_A, \ell_B \epsihat \}^n_\star \\
& & R^{n+1} = {i \over 4(n+1)} \theta^{AB} \left( \{\Omhat_A, (\ell_B + L_B)  \Rhat  \}^n_\star 
 - [\Rhat_{A},\Rhat_{B} ]^n_\star \right) \\
 & & \phi^{n+1}= {i \over 4(n+1)} \theta^{AB} \{\Omhat_A,  (\ell_B + L_B) \phihat \}^n_\star,
  ~~\dehat_{\hat \epsilon} \phihat =  \epsihat \star \phihat -  \phihat \star \epsihat \label{rec4}
\ena
where
$\Omhat_A$, $\Rhat_A$ are defined as the contraction $i_A$ along the tangent
vector $X_A$ of
the exterior forms $\Omhat$, $\Rhat$, i.e. $\Omhat_A\equiv i_A\Omhat$,
$\Rhat_A \equiv i_A \Rhat$. The apex $\,^n$ on the composite fields on
the right hand side
indicates that we are considering the term homogeneous of order
$\theta^n$ of the composite field.
We have also introduced the covariant Lie derivative $L_B$ along the
tangent vector $X_B$; it acts on $\Rhat$ and $\phihat$ as 
$L_B \Rhat =\ell_B \Rhat-\Omhat_B \star
\Rhat+ \Rhat \star\Omhat_B$ and
$L_B  \phihat = \ell_B \phihat - \Omhat_B \star \phihat +  \phihat \star \Omhat_B$.
In fact the covariant Lie derivative $L_B$ can be written in the Cartan form:
 \eq
  L_B = i_B D + D i_B~
    \en
where $D$ is the covariant derivative. The recursion formulae (\ref{rec1})-(\ref{rec4}) relate the $\theta^{n+1}$ order to the $\theta^n$ order of the NC fields. 

For the fields in the index-free geometrical action (\ref{scalartwisted}) the above formulae at first order become:
 \eqa
& & \varphi^{1}= {i \over 4} \theta^{AB} \{\Om_A,  (\ell_B + L_B) \varphi \} \label{varphi1} \\
& & \Phi^{1}= {i \over 4} \theta^{AB} \{\Om_A,  (\ell_B + L_B) \Phi \} \label{Phi1} \\
 & & V^1 = {i \over 4} \theta^{AB} \{\Om_A,  (\ell_B + L_B) V \} \label{V1} \\
 & & R^{1} = {i \over 4} \theta^{AB} \left( \{\Om_A, (\ell_B + L_B)  R  \}
 - [R_{A},R_{B} ] \right) \label{R1}
 \ena
 All these formulae are {\it not} $SO(1,3)$-gauge covariant, due to the presence of the ``naked" connection
$\Omhat$ and the non-covariant Lie derivative $\ell_A$. However, when
inserted in the NC action (\ref{scalartwisted}), the resulting action
is gauge invariant order by order in $\theta$. Indeed usual gauge variations
induce the $\star$-gauge variations under which the NC action is
invariant. Therefore the NC action, re-expressed in terms of ordinary
fields via the SW map, is invariant under usual gauge
transformations. Since these do not involve $\theta$, the expanded
action is invariant under ordinary gauge variations order by order in
$\theta$. This will be explicitly checked in the next sections for the first and second order $\theta$ correction of the NC action. 

\sect{Action at first order in $\theta$}

The first order correction of the Einstein term in (\ref{scalartwisted}) vanishes, as shown in ref. \cite{AC3}. The remaining two terms contribute to the first order correction $S^1$ of the action:
\eq
 S^1 = S^1_{\varphi\varphi VVVV} + S^1_{\varphi D\phi VVV}
 \en
\noi with
 \eqa
 & &\!\!\!\!\!\!\!\!\!\! S^1_{\varphi\varphi VVVV} = - {1 \over 4!} \int Tr \Big( i \ga_5 (( \varphi \star \varphi)^1 V\we  V \we V  \we V+
 \varphi \varphi (V \westar V \westar V \westar V)^1) \Big) \nonumber \\
& & =   {1 \over 48}  \theta^{AB} \int Tr \Big( \ga_5(( \unmezzo R_{AB} \varphi \varphi + \unmezzo \varphi\varphi R_{AB} 
+ L_A \varphi L_B \varphi ) V\we V \we V \we V+ \nonumber \\
& &~~~+ \varphi \varphi ~  ( L_A V \we L_B V \we V \we V + L_A (V \we V ) \we L_B (V \we V )+  V \we V \we L_A V \we L_B V)  ) \Big) \nonumber \\  \label{S1phiphi} 
 \ena
  The curvature components
$R_{AB}$ are defined as:
 \eq
 R_{AB} = i_B (R_A) =i_Bi_A R
 \en
 For the second term we need first to compute $(D \Phi)^1\equiv d (\Phi)^1 - (\Om \star  \Phi)^1 + (\Phi \star \Om)^1$. We find:
 \eq
 (D \Phi)^1 = {i \over 4} \theta^{AB}  (\{ \Om_A, (l_B + L_B) D \Phi \} - 2 \{R_A, L_B \Phi \})
 \en
Then
 \eqa
  & & S^1_{\varphi D\Phi VVV} = {1 \over 6} \int Tr \Big(i \ga_5( ( \varphi^1 D \Phi + \varphi (D \Phi)^1 +
  (D \Phi)^1 \varphi +  (D \Phi) \phi^1 + \nonumber \\
  & &+ {i \over 2} \theta^{AB} ( \ell_A \varphi \ell_B (D\Phi) + \ell_A (D \Phi) \ell_B \varphi) \we V \we V \we V +  (\varphi D\Phi + D \Phi \varphi) (V \westar V \westar V)^1) \Big)\nonumber \\ 
   \ena
   \noi Inserting the first order expressions for the fields yields 
\eqa
& & S^1_{\varphi D\Phi VVV} = - {1 \over 12} \theta^{AB} \int Tr \Big(  \ga_5  (( - \{ \varphi , \{ R_A, L_B \Phi \} \} +[ L_A \varphi, L_B(D \Phi) ]\nonumber \\
& & ~~~~~+ \unmezzo \{R_{AB}, \varphi D\Phi + D \Phi \varphi\}) \we V \we V \we V \nonumber\\
& & ~~~~~+ (\varphi D \Phi + D \Phi \varphi) \we (L_A V \we L_B (V \we V)+ V \we L_A V \we L_B V) ) \Big) \label{S1phidphi}
 \ena

Finally, inserting the expansions for the classical fields $R \equiv \unquarto R^{ab} \ga_{ab}$, 
 $\Phi^I \equiv \phi^I 1$, $\varphi^I \equiv \varphi^{Ia} \ga_a$, $V \equiv V^a \ga_a$, and carrying out the trace
 on spinor space yields a vanishing result for both contributions $S^1_{\varphi\varphi VVVV} $ and 
 $S^1_{\varphi D\Phi VVV}$. Indeed all terms have a gamma matrix content that does not contain the unit matrix, the only matrix with nonvanishing trace. Thus the first order (in $\theta$) correction to the classical action 
 (\ref{scalarsplusgravity}) vanishes, in agreement with the $\theta$-parity of the action discussed at the end of 
Section 3. To find nonzero contributions one has to compute the second order correction.

The expressions for the first order corrections (\ref{S1phiphi}) and
(\ref{S1phidphi}) are obtained algebraically via the geometrical
Seiberg-Witten map, that, together with the cyclicity of the integral,
ensures gauge invariance of the result. This is confirmed by the appearance
of only gauge-covariant terms in the integrands. Thus expressions (\ref{S1phiphi}) and (\ref{S1phidphi}) are a useful check, even if they eventually have to vanish  because of the
particular matrix structure of the index-free fields.

\sk
\noi {\bf Note:} one can also use the relations, due to the gamma matrix structure of the index-free classical fields:
 \eq
 \Om^T = C \Om C,~~V^T = C V C,~~\Phi^T=- C \Phi C,~~\varphi^T = C \varphi C,~~R^T = C R C \label{Crelations}
 \en
 \noi where $C$ is the charge conjugation matrix (see the Appendix). It is easy to prove $S^1_{\varphi\varphi VVVV}=S^1_{\varphi D\Phi VVV}=0$, simply by checking that, 
before taking the spinor trace, the transpose of the integrands in the action at first order are equal to minus the integrands. 

 \sect{Action at second order in $\theta$}

In this Section we expand the scalar action in  (\ref{scalarindexfree})  at second order
in $\theta$ using the recursive formulae of the geometric SW map  (\ref{rec1})-(\ref{rec4}).
In  \cite{ACD2} we have developed a method that allows to write each term in the expansion of a  generic noncommutative gauge theory action in explicit gauge invariant form. 
We apply this method to the scalar field action in (2.7) and obtain the second order corrections:
\eq
 S^2 = S^2_{\varphi\varphi VVVV} + S^2_{\varphi D\phi VVV}
 \en
 with (we omit wedge products):
  \eqa
 & &\!\!\!\!\!\!\!\!\!\! S^2_{\varphi\varphi VVVV}= { \theta^{AB} \theta^{CD}\over 4!\, 8} \int \!Tr \; i\ga_5 \Big( (\unmezzo \{ R_{CD},\unmezzo \{R_{AB},\varphi\varphi \} + 2 L_A \varphi L_B \varphi \} +\unmezzo  [L_C R_{AB},L_D(\varphi\varphi)]  \nonumber \\[.3em]
& & ~~~~~ - \unmezzo  \{ \{R_{AC},R_{BD} \} , \varphi\varphi \} + (L_AL_C \varphi)(L_BL_D \varphi) -  [ \{R_{AC} , L_B \varphi \} , L_D \varphi ] ) VVVV \nonumber \\[.3em]
& & ~~~~~ +  ( \{R_{CD}, \varphi\varphi \} +2  L_C \varphi L_D \varphi) ( \{L_AV L_B V, VV \}+ L_A(VV)L_B(VV) ) \nonumber \\[.3em]
& & ~~~~~~ + \varphi \varphi (-  \{[\{R_{AC},L_BV \},L_DV],VV \} -  [ \{ \{R_{AC},L_BV \}, V\} ,L_D(VV) ] \nonumber \\[.3em]
 & & ~~~~~ +  [L_C(L_AV L_BV),L_D(VV)] + \{(L_AL_CV)(L_BL_DV), VV\} + 2 L_AVL_BVL_CVL_DV \nonumber \\[.3em]
& & ~~~~~ + [[ L_AL_CV,L_BV],L_D(VV)] + L_AL_C(VV) L_BL_D(VV) ) \Big)
 \ena
 \noi and
 \eqa
& &\!\!\!\!\!\!\!\!\!\! S^2_{\varphi D\phi VVV} = - {\theta^{AB}
 \theta^{CD}\over 6 \cdot 8} \!\!\int\! Tr \;i \ga_5 \Big(
 {1 \over 4}  \{ R_{CD}, \{ R_{AB}, \{ \varphi , D \phi \} \} +2 [L_A \varphi , L_B D \phi] - 
 2 \{ \varphi, \{ R_A , L_B \phi \} \} \} 
 \nonumber \\[.3em]
    & & + \unmezzo [L_C R_{AB} , L_D \{\varphi, D \phi \} ]  - \unmezzo \{ \{ R_{AC},R_{BD} \}, \{ \varphi, D\phi \} \} - 2 [ L_A \phi , L_B \{R_C,L_D \phi \} ]
\nonumber \\[.3em]
 & & -  [ \{ R_{CA},L_D \varphi \} , L_B D \phi] -  [L_A \varphi , \{R_{CB} , L_D D \phi \} ] + \{L_CL_A \varphi, L_D (L_B D \phi) \} \nonumber \\[.3em]
& & 
- \{ \varphi , [L_C R_A , L_DL_B \phi ] - \{ i_A (R_CR_D), L_B \phi \} -  \{ R_A, \{R_{CB} , L_D \phi \} \} \}
 \nonumber \\[.3em]
& & +  ( [R_{CD}, \{ \varphi, D \phi \}] +  2 [L_C \varphi , L_D D \phi] - 2 \{ \varphi , \{ R_C , L_D \phi \} \} )  ((L_A V) L_B (VV) + V (L_A V) (L_B V)) \nonumber \\[.3em]
    & &+ \{ \varphi, D \phi \} ( (L_C L_A V) L_D L_B (VV) + L_A V [L_C L_B V , L_D V ] + V (L_C L_A V) (L_D L_B V)  \nonumber \\[.3em]
  & & + (L_C V) L_D (L_A V L_B V) -  \{R_{CA} , L_D V \} L_B (VV)  \nonumber \\[.3em]
 & &  -  L_A V \{ V, \{ R_{CB} , L_D V \} \} - V [ \{ R_{CA} , L_D V \} , L_B V ] \Big) 
 \ena
\noi {\bf Note: } by taking $\varphi = const.$ in $S^2_{\varphi\varphi VVVV}$ one obtains the explicit second order correction to the cosmological term (simply by discarding all terms containing derivatives of $\varphi$).

 \sect{Dynamical noncommutativity}

The background commuting vector fields $X_A$, defining the twist, can
become dynamical if we relate them to the scalar fields $\phi^I$.

Consider a spacetime manifold $M$ that
can be described by a single coordinate system, with coordinates $x^\mu$. In this case the
vector fields $X_A$ (that are globally defined on $M$) can be written
$X_A=X^\mu_A\partial_\mu$.  They are then  identified with the inverse of the matrix given by the derivatives of a ``potential" $Z^A(\phi)$:
  \eq
  X_A^\mu \equiv (\part_\mu Z^A)^{-1}\label{dynX}
  \en
  where the index $I$ labelling the scalars $\phi^I$ is now chosen to coincide with the index $A$ labelling the vector fields, and $A$ runs on 1,2,3,4. 
  This definition automatically ensures that the vector fields commute, i.e. that
 \eq
 [X_A,X_B] =0 \,\Leftrightarrow \,X_A^\mu \part_\mu X_B^\nu - X_B^\mu \part_\mu X_A^\nu =0~.
 \en
For example, we can choose $Z^A = \phi^A$, as in ref. \cite{ACD}. One has then to check whether there exist solutions of the scalar field equations and the Einstein equations for which $(\part_\mu \phi^A)$ is
invertible. The study of these solutions, and the analysis of their stability, is postponed to future work.
If a solution exists with  $\part_\mu \phi^A = \de^A_\mu$ ($\phi^A = x^A+const$), it would correspond to $X_A^\mu=\de^\mu_A$, i.e. to
Moyal noncommutativity. 

\noi Another choice for $Z$ is:
 \eq
 Z^A = {\phi^A \over \phi^2},~~\phi^2 \equiv \phi^B \phi^B
 \en
 leading to
 \eq
  (X_A^\mu)^{-1} = \part_\mu ({\phi^A \over \phi^2}) = {\part_\mu \phi^B \over \phi^2} (\de^{AB} -  2 {\phi^A \phi^B \over \phi^2})
 \en
 \noi or:
  \eq
X_A^\mu =(\part_\mu \phi^B)^{-1}\phi^2 (\de^{AB} -  2 {\phi^A \phi^B \over \phi^2})
 \en
\noi since $(\de^{AB} -  2 {\phi^A \phi^B \over \phi^2})$ is its own inverse.
 In this case a solution $\phi^A = x^A+const$ for $const =0$ would correspond to
 \eq
X_A^\mu = \de_A^\mu x^2 - 2 x^A x^\mu  
 \en
 \noi so that spacetime becomes commutative close to the origin of coordinates.

\appendix

\sect{Gamma matrices in $D=4$}

We summarize in this Appendix our gamma matrix conventions in $D=4$.

\eqa
& & \eta_{ab} =(1,-1,-1,-1),~~~\{\ga_a,\ga_b\}=2 \eta_{ab},~~~[\ga_a,\ga_b]=2 \ga_{ab}, \\
& & \ga_5 \equiv i \ga_0\ga_1\ga_2\ga_3,~~~\ga_5 \ga_5 = 1,~~~\epsi_{0123} = - \epsi^{0123}=1, \\
& & \ga_a^\dagger = \ga_0 \ga_a \ga_0, ~~~\ga_5^\dagger = \ga_5 \\
& & \ga_a^T = - C \ga_a C^{-1},~~~\ga_5^T = C \ga_5 C^{-1}, ~~~C^2 =-1,~~~C^\dagger=C^T =-C
\ena

\subsection{Useful identities}

\eqa
 & &\ga_a\ga_b= \ga_{ab}+\eta_{ab}\\
 & & \ga_{ab} \ga_5 = {i \over 2} \epsilon_{abcd} \ga^{cd}\\
 & &\ga_{ab} \ga_c=\eta_{bc} \ga_a - \eta_{ac} \ga_b -i \epsi_{abcd}\ga_5 \ga^d\\
 & &\ga_c \ga_{ab} = \eta_{ac} \ga_b - \eta_{bc} \ga_a -i \epsi_{abcd}\ga_5 \ga^d\\
 & &\ga_a\ga_b\ga_c= \eta_{ab}\ga_c + \eta_{bc} \ga_a - \eta_{ac} \ga_b -i \epsi_{abcd}\ga_5 \ga^d\\
 & &\ga^{ab} \ga_{cd} = -i \epsi^{ab}_{~~cd}\ga_5 - 4 \de^{[a}_{[c} \ga^{b]}_{~~d]} - 2 \de^{ab}_{cd}\\
& & Tr(\ga_a \ga^{bc} \ga_d)= 8~ \de^{bc}_{ad} \\
& & Tr(\ga_5 \ga_a \ga_{bc} \ga_d) = -4 \;\!i \epsi_{abcd} \\
& & Tr(\ga^{rs} \ga_a \ga_{bc} \ga_d)=4(-2 \de^{rs}_{cd} \eta_{ab} + 2 \de^{rs}_{bd} \eta_{ac} - 3! \de^{rse}_{abc} \eta_{ed}) \\
& & Tr(\ga_5 \ga^{rs} \ga_a \ga_{bc} \ga_d)=
4(-i \eta_{ab} \epsi^{rs}_{~~cd} + i \eta_{ac} \epsi^{rs}_{~~bd} + 2i \epsi_{abc}^{~~~e} \de^{rs}_{ed})
 \ena
\sk
\noi where
$\delta^{ab}_{cd} \equiv \frac{1}{2}(\delta^a_c\delta^b_d-\delta^b_c\delta^a_d)$, $\delta^{rse}_{abc} \equiv  {1 \over 3!} (\de^r_a \de^s_b \de^e_c$ + 5 terms), 
and indices antisymmetrization in square brackets has total weight $1$.

\end{document}